\documentclass[%
 reprint,
nofootinbib,
 amsmath,amssymb,
 aps,
onecolumn]{revtex4-2}

\usepackage{graphicx}
\usepackage{dcolumn}
\usepackage{bm}
\usepackage[]{hyperref}
\usepackage{amsmath}
\usepackage{mathrsfs}
\usepackage{gensymb}
\usepackage{tensor}
\usepackage{physics}
\usepackage{amsfonts}
\usepackage{amssymb}
\usepackage{comment}
\usepackage{subfigure}
\usepackage{xcolor}
\usepackage{soul}
\usepackage{lipsum}
\usepackage{csquotes}
\usepackage{float}
\usepackage[normalem]{ulem}
\usepackage{amssymb}
\begin{document}
\title{Spontaneous scalarization around a black hole in a dark matter halo}
\author{ M. Navid Gasemi Zad \footnote{meysam.navid@ut.ac.ir}, H. Mohseni Sadjadi\footnote{mohsenisad@ut.ac.ir}
\\ {\small Department of Physics, University of Tehran}}
\begin{abstract}
We demonstrate that in a scalar-tensor theory where dark matter is conformally coupled to a scalar field, a black hole surrounded by a Hernquist dark matter halo can undergo spontaneous scalarization, acquiring a nontrivial scalar field configuration.
The astrophysical properties of the dark matter halo directly dictate the quantum-like spectrum of scalar field configurations around the central black hole. Remarkably, the conformal coupling constant becomes discretized and is determined solely by the halo's compactness parameter.
\end{abstract}
\maketitle

\section{Introduction}
The nature of the strong-gravity regime has been studied extensively. A hot topic in this context is studying the possibility of the existence of a nontrivial scalar field in the background of a black hole (BH). In \cite{Chase:1970omy}, it was shown that a regular massless scalar field cannot exist in the background of a static asymptotically flat black hole. In \cite{Bekenstein:1972ny},  it was shown that canonical minimally coupled scalar fields do not exist in the exterior of a stationary black hole background. The role of the scalar field in gravity and cosmology has motivated attempts to modify general relativity (GR), such as adding higher curvature terms to the action or introducing new degrees of freedom, such as a scalar field and coupling it to different curvature terms(non-minimal coupling). A broad range of modified gravity models has been developed \cite{C. Brans, T. Clifton, T. P. Sotiriou, L. Heisenberg, E. N. Saridakis, C. Charmousis}. One of the main frameworks for modifying gravity is the scalar-tensor (ST) theories that introduce a dynamical scalar field alongside the tensor field of GR.
The first example, Brans-Dicke theory, was formulated to incorporate Mach’s principle by allowing the gravitational constant to vary \cite{C. Brans}. Other scalar-tensor models are constructed to reduce to GR in appropriate limits (ensuring consistency with solar-system experiments) while offering richer behavior in large-scale structure or in other extreme conditions. The most general form of these theories is Horndeski theory \cite{G. W. Horndeski, C. Deffayet, T. Kobayashi, C. Deffayet2, T. Kobayashi2}. In some of these models, \cite{J. D. Bekenstein1,J. D. Bekenstein2,N. M. Bocharova}, the scalar field is not regular at the horizon. However certain scalar-tensor theories allow the black hole to support a nontrivial scalar field configuration. For instance, when the scalar is coupled to additional invariants like the Gauss-Bonnet (GB) term \cite{P. Kanti}, \cite{G. Antoniou1}, \cite{G. Antoniou2}. In this outstanding model, a special phenomenon named spontaneous scalarization makes the scalar field nontrivial in the background of strongly gravitating objects like BHs\cite{Esposito-Farese:1993gds, Damour:1996ke, Damour:1996ke, Doneva:2017bvd,Silva:2017uqg,sajj}.This phenomenon occurs when the squared effective mass takes negative values, and makes the BH solutions of ST theories different from their GR counterparts. Einstein scalar-Gauss-Bonnet (EsGB) theories provide a promising arena to explore strong-gravity phenomena and to confront theoretical predictions with observational data. Some cosmological aspects of GB gravity as well as neutrino oscillations in this context, are discussed by \cite{MohseniSadjadi:2025ztu, MohseniSadjadi:2023amn}. Apart from the GB term, the presence of matter can also lead to scalarization\cite{Cardoso:2013fwa, Cardoso:2013opa},\cite{SZ}. In fact, black holes are not alone; they may be surrounded by matter in the form of an accretion disk or by a galaxy. There are evidences, such as the rotation curve of spiral galaxies \cite{Sofue:2000jx}, the dynamics of galaxy clusters \cite{Frenk:1995fa}, the cosmic microwave background radiation \cite{Gawiser:2000az}, and the extremely high mass-luminosity ratios of elliptical galaxies, demonstrating that dark matter exists as haloes in galaxies.

In this work, we investigate spontaneous scalarization via a Hernquist dark matter halo. We study the condition for the existence of a non-trivial scalar field in the black hole-dark matter (BH-DM) spacetime when it is conformally coupled to a halo around a central BH. We show that the astrophysical concept known as the halo's compactness characterizes the DM halo-induced scalarization.

We use units $\hbar=c=G=1$

\section{Black hole embedded in a Hernquist dark matter halo, and spontaneous scalarization}
\label{SS}
Spontaneous scalarization was first discussed in\cite{Esposito-Farese:1993gds} and then in\cite{Damour:1996ke}. In scalar-tensor theories of gravity, when a scalar field couples to matter via conformal coupling, a neutron star (NS) can acquire a nontrivial scalar field for some range of coupling constants through tachyonic instability. Neutron stars then have properties that can be dramatically different from their GR counterparts. This was called spontaneous scalarization.
A similar mechanism operates in black holes when the scalar field is coupled to curvature invariants, such as in the scalar-Gauss-Bonnet (sGB).
In sGB model, the Gauss-Bonnet (GB) invariant couples to a specific function of scalar fields. This type of coupling shows spontaneous scalarization\cite {Doneva:2017bvd,Silva:2017uqg}. The corresponding action is
\begin{align}
\label{sGB}
\mathcal{S}=\frac{1}{2}\int d^4x\sqrt{-g}[R-\frac{1}{2}\nabla_\mu\Phi\nabla^\mu\Phi+f(\Phi)\mathcal{G}],
\end{align}
where
\begin{align}
\label{GB inva}
\mathcal{G}= R^2 - R_{\mu\nu}R^{\mu\nu} + R_{\mu\nu\rho\sigma} R^{\mu\nu\rho\sigma},
\end{align}
is GB invariant.
By variation of the action above with respect to $\Phi$,
\begin{align}
\label{sGB KG}
\square\Phi+f_{,\Phi}(\Phi)\mathcal{G}=0,\,\,\,\,f_{,\Phi}(\Phi)=df/d\Phi.
\end{align}
There is a trivial solution $\Phi=\Phi_0$ provided that $(df/d\Phi)_{\Phi=\Phi_0}=0$.
In the linear approximation
\begin{align}
\label{curved}
\square\delta\Phi+\mu^2_{eff}\delta\Phi=0.
\end{align}
If $\mu^2_{eff}=-f_{,,\Phi}(\Phi_0)\mathcal{G}<0$ then there is a tachyonic instability. In\cite{Silva:2017uqg} it was shown that for coupling function $f=\eta\frac{\Phi^2}{8}$ which satisfies $(df/d\Phi)_{\Phi=\Phi_0}=0$ the scalar field has a regular solution in black hole spacetime. For a discrete spectrum of the coupling parameter $\eta$, there are nontrivial regular solutions.
In the following we focus on a similar formalism but with a different source of scalarization: the coupling of the scalar field to the dark matter halo.

\subsection{Black hole embedded in a Hernquist dark matter halo}
Nowadays, we know that there is a supermassive black hole at the center of a galaxy and that the black hole is surrounded by dark matter halo whose distribution can be  obtained from numerical simulations. There are several profiles for galactic dark matter haloes. One of them is the Hernquist profile\cite{6}
\begin{align}
\label{Hern}
\rho(r)=\frac{m}{2\pi}\frac{a}{r(r+a)^3},
\end{align}
where $a$ is the scale radius of the halo , and $m$ is the total mass of the halo. We consider the galactic DM halo to be spherically symmetric
\begin{align}
\label{halo}
ds^2=-f(r)dt^2+f(r)^{-1}dr^2+r^2d\theta+r^2\sin^2\theta d\phi^2.
\end{align}
The mass of the halo is
\begin{align}
\label{mass of halo}
m(r)=4\pi\int_0^r\rho(r^\prime)r^{\prime2}dr^\prime.
\end{align}
The tangential velocity $v^2_{\text{tg}}=m(r)/r$  of a test particle moving in the
dark halo is\cite{Matos:2003nb, Xu:2018wow, Jusufi:2019nrn, Al-Badawi:2024asn, Gohain:2024eer}
\begin{align}
\label{v from f}
v^2_{tg}=\frac{r}{\sqrt{f(r)}}\frac{d\sqrt{f(r)}}{dr}=\frac{r(d\ln\sqrt{f(r)})}{dr}.
\end{align}
So we obtain
\begin{align}
\label{Hern1}
f(r)=\exp(2m/a)\exp(-2m/(r+a)).
\end{align}
The energy-momentum tensor of dark matter is ${T_m}\,^\mu \,_\nu= \text{diag}[-\rho,-\rho,p_t,p_t]$.
Then, by solving Einstein's equation for the DM halo, we have\cite{Nieto:2025apz, Ali:2025ney, Jha:2025xjf, Jumaniyozov:2025xxh, Ahmed:2025ttq, Senjaya:2025via, Uktamov:2025lwb}
\begin{align}
\label{pressure}
p_t=[\frac{ma^2}{2\pi}\frac{1}{r(r+a)^4}+\frac{m/a}{2\pi}\frac{a(a+m)}{(r+a)^4}]\exp(2m/a)\exp(-2m/(r+a)).
\end{align}

\begin{figure}[H]
\includegraphics[width=3in]{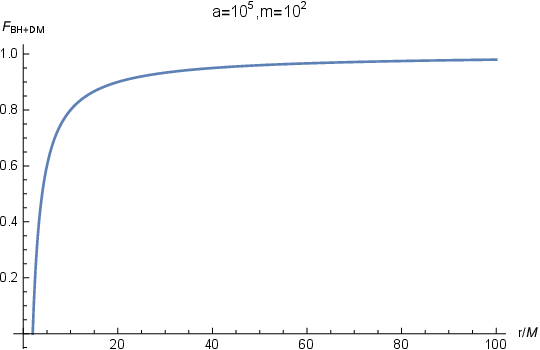}
\caption{Function $F$ of the composed BH and DM for $a=10^5, m=10^2$}.
\label{fig1}
\end{figure}

We consider a spacetime consisting of a BH embedded in the above DM halo so that
\begin{align}
\label{BH+halo}
ds^2=-(f(r)+F_1(r))dt^2+(f(r)+F_1(r))^{-1}dr^2+r^2d\theta+r^2\sin^2\theta d\phi^2.
\end{align}
The Einstein field equation can now be written as
\begin{align}
\label{E equ}
G^\nu\,_\mu=8\pi (T^\nu\,_\mu(BH)+ {T_m}\,^\mu \,_\nu).
\end{align}
By solving this equation for the above metric we have\cite{Nieto:2025apz, Ali:2025ney, Jha:2025xjf, Jumaniyozov:2025xxh, Ahmed:2025ttq, Senjaya:2025via, Uktamov:2025lwb}
\begin{align}
\label{met fun}
F(r)\equiv F_1(r)+f(r)=-\frac{2M}{r}+\exp(2m/a)\exp(-2m/(r+a)).
\end{align}
The spacetime is asymptotically flat, and for $\frac{m}{a}\ll 1$, the event horizon is approximately located at $2M$ (see FIG.(\ref{fig1})).

There is a hierarchy of scales, i.e., $M\ll m\ll a$, which ensures that there are no curvature singularities outside
the horizon\cite{Pezzella:2024tkf}. The compactness parameter $\mathcal{C}=\frac{m}{a}$, helps to compare different halo configurations. For galactic haloes  $\mathcal{C}\ll 1$.

\subsection{Spontaneous scalarization}
\begin{figure}[H]
\includegraphics[width=3in]{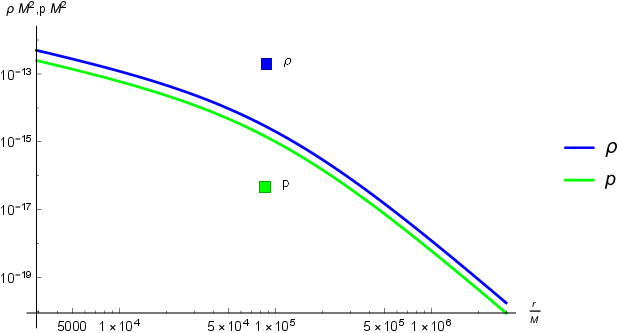}
\caption{Density(blue) and pressure(green) of the DM halo for $a=10^5, m=10^2$}
\label{figrhop}
\end{figure}

As discussed in Section (\ref{SS}), in some scalar-tensor theories of gravity, the scalar field can experience a negative
effective mass squared called a tachyonic mode.
We work with the following model
\begin{align}
\label{Sadjadi}
\mathcal{S}=\int d^4x\sqrt{-g}[\frac{R}{16\pi}-\frac{1}{2}\nabla_\mu\Phi\nabla^\mu\Phi]+\mathcal{S}_m[\Psi_m;\mathcal{A}^2(\Phi)g_{\mu\nu}],
\end{align}
where DM couples conformally to the geometry via the scalar field.
The action is written in the Einstein frame, which means that the scalar field is minimally coupled to gravity and has a canonical kinetic term. The coupling with the matter field is through the function $\mathcal{A}(\Phi)$.
We choose
\begin{align}
\label{conf fun}
\mathcal{A}(\Phi)=\exp\left(\frac{\eta}{2}\Phi^2\right),
\end{align}
where $\eta$ is a real constant which we will see is not positive.
The equation of motion of the scalar field is
\begin{align}
\label{Khoda}
\square\Phi=-(\ln\mathcal{A})_{,\Phi}T_m
\end{align}
where $T_m=T_m^{\mu\nu}g_{\mu\nu}$ is the trace of the energy-momentum tensor of dark matter.
For the DM halo $T_m=-2\rho+2p_t$. For convenience at the rest of the paper we omit the index "t". So
\begin{align}
\label{not KG}
\square\Phi=2\frac{\mathcal{A}_{,\Phi}}{\mathcal{A}}(\rho-p).
\end{align}
By multiplying the above equation with $\frac{\mathcal{A}_{,\Phi}}{\mathcal{A}}$ and integration over spacetime (volume V), we obtain
\begin{align}
\label{int 1}
\int_V (\frac{\mathcal{A}_{,\Phi}}{\mathcal{A}}\nabla_\mu\nabla^\mu\Phi-2(\frac{\mathcal{A}_{,\Phi}}{\mathcal{A}})^2(\rho-p))\sqrt{-g}d^4x=0.
\end{align}
Then integration by part gives
\begin{align}
\label{int 2}
\int_{\partial V}(\frac{\mathcal{A}_{,\Phi}}{\mathcal{A}}\nabla_\mu\Phi)n^\mu\sqrt{h}d^3x-\int_V((\frac{\mathcal{A}_{,\Phi}}
{\mathcal{A}})_{,\Phi}\nabla_\mu\Phi\nabla^\mu\Phi+2(\frac{\mathcal{A}_{,\Phi}}{\mathcal{A}})^2(\rho-p))\sqrt{-g}d^4x=0,
\end{align}
where $\partial V$ is the boundary of $V$ and $n^\mu$ is the normal to the
boundary. $V$ is bounded by the BH horizon, two partial Cauchy surfaces, and spatial infinity. The contribution of the boundary term vanishes. The horizon contribution vanishes by symmetry, as the normal to the horizon is a Killing vector(the horizon is a Killing horizon). The contribution of the boundary at infinity vanishes because of asymptotic flatness. The contributions of the Cauchy surfaces exactly cancel each other, as they can be generated by an isometry.
\begin{align}
\label{int 3}
\int_V((\frac{\mathcal{A}_{,\Phi}}{\mathcal{A}})_{,\Phi}\nabla_\mu\Phi\nabla^\mu\Phi+2(\frac{\mathcal{A}_{,\Phi}}
{\mathcal{A}})^2(\rho-p))\sqrt{-g}d^4x=0.
\end{align}
The necessary condition for spontaneous scalarization is then\cite{9}
\begin{align}
\label{int 2}
\mu^2_{eff}=2(\rho-p)(\frac{\mathcal{A}_{,\Phi}}{\mathcal{A}})_{,\Phi}<0
\end{align}
By choosing the conformal function (\ref{conf fun}), the equation of motion (E.O.M), of the scalar fields is
\begin{align}
\label{KG 2}
\square\Phi=2\eta(\rho-p)\Phi.
\end{align}
In the following we consider $\rho-p>0$. According to (\ref{Hern}), and (\ref{pressure}), this inequality is satisfied when, $r<\frac{a^2}{m}=\frac{a}{\mathcal{C}}$. As an illustration in Fig. (\ref{figrhop}), we have plotted $p$ and $\rho$ for $a=10^5, m=10^2$.
Let us discuss this inequality for a gravitationally bound galactic halo described by the Hernquist profile. By substituting the virial radius
$ r=r_{vir}$, and defining the concentration parameter for this profile as $c=\frac{r_{vir}}{a}$ \cite{Cuesta:2007it}, we obtain the condition $c<\frac{1}{\mathcal{C}}$. For the parameters used in our illustration we have $\frac{1}{\mathcal{C}}=10^3$, which is comfortably
larger than the typical concentration parameters of astrophysical halos ($c\sim 3-10$).
Therefore by assuming that the trace of dark matter energy momentum tensor is negative, the condition for tachyonic instability becomes $\eta<0$.

By the decomposition:
\begin{align}
\label{var sep}
\Phi(t,r,\theta,\phi)=\sum_{l=0}^{\infty}\sum_{m=-l}^l\frac{\psi}{r}Y_{lm}(\theta,\phi)e^{iwt},
\end{align}
we obtain a Schrödinger-like equation for the radial section
\begin{align}
\label{Schrodinger-like}
\frac{d^2\psi}{dr_*^2}-V\psi=0,
\end{align}
where $\frac{dr}{dr_*}=F(r)$ is tortoise coordinates.
The effective potential is
\begin{align}
\label{eff pot}
V(r)=F(r)\left[\frac{1}{r}\frac{dF}{dr}+\frac{l(l+1)}{r^2}-2\abs{\eta}(\rho-p)_{DM}\right].
\end{align}
From now on, we set $l=0$  (spherically symmetric solution).
The effective potential is plotted below for  $a=10^5, m=10^2$. A large negative value of $\eta$ provides a deep negative potential.

\begin{figure}[H]
\includegraphics[width=4in]{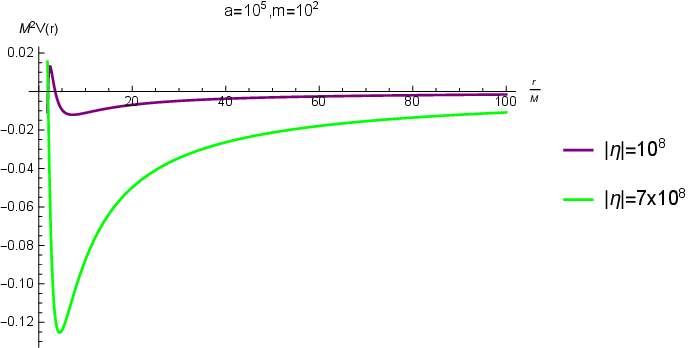}
\caption{Effective potentials for $a=10^5, m=10^2$}
\label{Arm}
\end{figure}

\subsection{WKB approximation and quantization condition}
The Schrodinger-like radial differential equation (\ref{Schrodinger-like}) has a mathematical form that is amenable to a standard WKB analysis if the WKB validity condition$\abs{V^\prime}\ll  \abs{V^{3/2}}$ is satisfied. By evaluating this condition at the virial radius $ r=r_{vir}$
and using the halo parameters with $c=\frac{r_{vir}}{a}$ we obtain the requirement:
\begin{align}
\label{validity}
\frac{M^2\mathcal{C}^2}{a^2(1+c)^6\sqrt{2\pi}}\sqrt\abs{\eta}\gg 1.
\end{align}
This condition guarantees that the coupling constant $\abs{\eta}$ is large and places us in the eikonal regime.
The quantization condition is given by the Bohr-Sommerfeld rule \cite{11}:
\begin{align}
\label{WKB}
\int_{r_*^-}^{r_*^+}\sqrt{-V(r_*;\eta)}dr_*=(n+\frac{1}{2})\pi
\end{align}
$r_*^\pm$ are classical turning points where the effective potential vanishes, and the resonant parameter $n$ is a a non-negative integer\cite{12}. The zeroes of the effective potential are derived from
\begin{align}
\label{turning points}
F(r_h)=0,\,\,\,\,\,Y(r) \equiv\left[\frac{1}{r}\frac{dF}{dr}-2\abs{\eta}(\rho-p)_{DM}\right]_{r_2^>}=0
\end{align}
In the astrophysically relevant limit $\frac{m}{a}\ll 1$, and using the Hernquist density profile (which behaves like $r^{-1}$ for $r\ll a$, and $r^{-4}$ for $(r\gg a)$), (\ref{turning points}) yields four roots:
\begin{align}
\label{4roots}
r_1^{\lessgtr}=\mp2a\sqrt{\frac{M\pi}{m\abs{\eta}}}\lessgtr0 \,\,\,\,\,\,\,,and\,\,\,\,\,\,\,r_2^{\lessgtr}=\frac{(a-m)m\abs{\eta}\mp\sqrt{m\abs{\eta}(-16a^2M\pi+(a-m)^2m\abs{\eta})}}{8M\pi}
\end{align}
The roots $r_1^{\lessgtr}$ are symmetrical about the origin, and the relevant positive root is $r_1^>$. The WKB validity condition (\ref{validity}) ensures that this root is well inside the event horizon:
\begin{align}
\label{small root}
r_1^>=2a\sqrt{\frac{M\pi}{m\abs{\eta}}}< r_h
\end{align}
Therefore it is not a physical turning point in the exterior spacetime.
For the roots $r_2^{\lessgtr}$, (\ref{validity}) allows to make the approximation $(-16a^2M\pi+(a-m)^2m\abs{\eta})\simeq (a-m)^2m\abs{\eta}$ since
$16a^2M\pi\ll (a-m)^2m\abs{\eta}\simeq a^2m\abs{\eta}$. This simplifies the positive root $r_2^>$ to
\begin{align}
\label{big root}
r^+=r_2^>=\frac{(a-m)m\abs{\eta}+\sqrt{m\abs{\eta}(-16a^2M\pi+(a-m)^2m\abs{\eta})}}{8M\pi}\simeq\frac{2(a-m)m\abs{\eta}}
{8M\pi}
\rightarrow\infty.
\end{align}
Thus, in this large $\abs{\eta}$ limit, the second turning point is located at spatial infinity.
In summary, the analysis of the roots, supported by the WKB validity condition, identifies the two classical turning points of the effective potential as: $r^-=r_h$ and $r^+=r_2^>$.
Consequently, the potential well that can support bound-state solutions for the scalar field is bounded by the event horizon and extends to infinity (see Fig. (\ref{Arm})).

With these turning points, the quantization condition (\ref{WKB}) becomes:
\begin{align}
\label{appro}
\int_{r^-}^{r^+}\sqrt{\abs{\frac{\left[\frac{1}{r}\frac{dF}{dr}-2\abs{\eta}(\rho-p)\right]}{F(r)}}}dr=(n+\frac{1}{2})\pi
\end{align}

\subsection{Analytic treatment}
Eq.(\ref{validity}) implies that the second term in $V(r)$ is dominant (large coupling constant) which means relation(\ref{appro}) is:
\begin{align}
\label{analytic}
\int_{r_*^-}^{r_*^+}\sqrt{-V(r_*;\eta)}dr_*=X(m,a)\sqrt{\abs{\eta}}
\end{align}
Where
\begin{align}
\label{X}
X(m,a)=\left[\frac{(a+r) \sqrt{\frac{m \left(a^2+a r-m r\right)}{(a+r)^4 (r-2 M)}} \left(\frac{(a+r) \sqrt{r-2 M}
   \left(a^2+2 a M-2 m M\right) \tan ^{-1}\left(\frac{\sqrt{a m} \sqrt{r-2 M}}{\sqrt{a+2 M}
   \sqrt{a^2+a r-m r}}\right)}{\sqrt{a m} (a+2 M)^{3/2} \sqrt{a^2+a r-m r}}+\frac{r-2 M}{a+2
   M}\right)}{\sqrt{2 \pi }}\right]_{r^-}^{r^+}
\end{align}
 In the astrophysical regime, i.e.  $M\ll m\ll a$
\begin{align}
\label{double check analytic treatment}
X(m,a)=\frac{1}{\sqrt{2\pi}}\left[\sqrt{\frac{m}{a}}\sqrt{\frac{r-2M}{r+a}}+\arctan(\sqrt{\frac{m}{a}}
\sqrt{\frac{r-2M}{r+a}})\right]_{r^-}^{r^+}
\end{align}
Finally
\begin{eqnarray}
\label{analytic treatment}
X(m,a)&=&\frac{1}{\sqrt{2\pi}}\left[\sqrt{\frac{m}{a}}+\arctan(\sqrt{\frac{m}{a}})\right]\nonumber \\
&\approx&\sqrt{\frac{2}{\pi}}\sqrt{\mathcal{C}}
\end{eqnarray}
From (\ref{WKB}) and (\ref{analytic}) we have:
\begin{equation}
\eta_n=-\frac{(n+\frac{1}{2})^2\pi^3}{2\mathcal{C}},
\end{equation}
showing that the coupling is quantized in terms of the inverse of the halo compactness in the Hernquist model. We note that this
is more accurate in the large coupling regime $n\gg 1$ in the eikonal regime.

By defining
\begin{equation}\label{diff of sqrts}
\Delta_n :=\sqrt{\abs{\eta_{n+1}}}- \sqrt{\abs{\eta_n}},
\end{equation}
we obtain
\begin{align}
\label{treatment}
\Delta_n X(m,a)=\pi
\end{align}
In the following we study numerically the integral (\ref{appro}), and investigate the validity of (\ref{treatment}).

\subsection{Numerical investigation}
The first three quantized values of $\eta$'s is derived numerically from  (\ref{WKB}), (\ref{appro}), for $a=10^5, m=10^2 (\mathcal{C}=10^{-3})$, and reported in TAB. \ref{etas1}.
\begin{table}[H]
\centering
\begin{tabular}{|c|c|c|c|} 
 \hline n &  0 & 1 & 2  \\ \hline
$\abs{\eta_n}$ & 3316.5058446458006& 40324.339974315604 & 103308.4908812381\\ \hline
\end{tabular}
\caption{Discrete couplings for $a=10^5, m=10^2$}
\label{etas1}
\end{table}

\begin{figure}[H]
\includegraphics[width=5.5in]{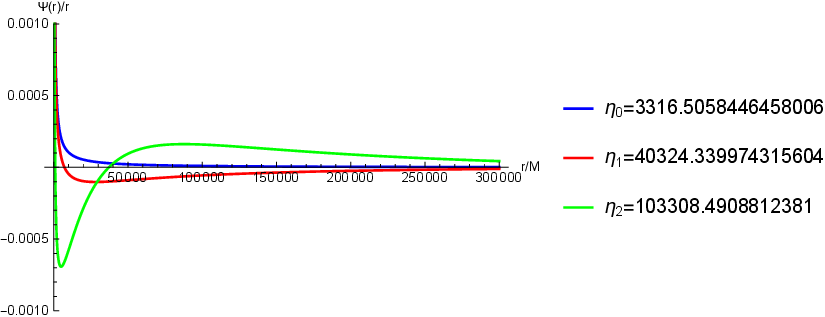}
\caption{Scalar field and its nodes for different couplings for $a=10^5, m=10^2$}
\label{fig2}
\end{figure}

We plot $\frac{\psi}{r}$ from (\ref{Schrodinger-like}) for  $a=10^5, m=10^2, (\mathcal{C}=10^{-3})$ in figure (\ref{fig2})
For discrete values of coupling parameter $\eta_n$ , there are $n$ nodes in the radial profile of the scalar field.

By solving (\ref{appro}), the first quantized coupling constants for halo parameters $a=10^6, m=10^3 (\mathcal{C}=10^{-3})$, and
 $a=10^7, m=10^2, (\mathcal{C}=10^{-5})$ are reported in TAB.\ref{etas for 6,3}, and TAB.\ref{etas for seven, two} respectively.
\begin{table}[H]
\centering
\begin{tabular}{|c|c|c|c|c|} 
 \hline n &  0 & 1 & 2 & 3 \\ \hline
$\abs{\eta_n}$ &4409.062745423526& 39777.07517255942& 114007.46838280425&205671.75013641946\\ \hline
\end{tabular}
\caption{Discrete couplings for $a=10^6, m=10^3$}
\label{etas for 6,3}
\end{table}

\begin{table}[H]
\centering
\begin{tabular}{|c|c|c|c|c|} 
 \hline n &  0 & 1 & 2 & 3 \\ \hline
$\abs{\eta_n}$ &410968.2101856222& 3605727.861429465& 9940320.335028399&1940619.3898644842\\ \hline
\end{tabular}
\caption{Discrete couplings for $a=10^7, m=10^2$}
\label{etas for seven, two}
\end{table}
As we see in TABs.\ref{etas for 6,3} , and (\ref{etas for seven, two}), for having fixed node number, as the compactness parameter decreases, the coupling increases.

\begin{figure}[H]
\includegraphics[width=4.5in]{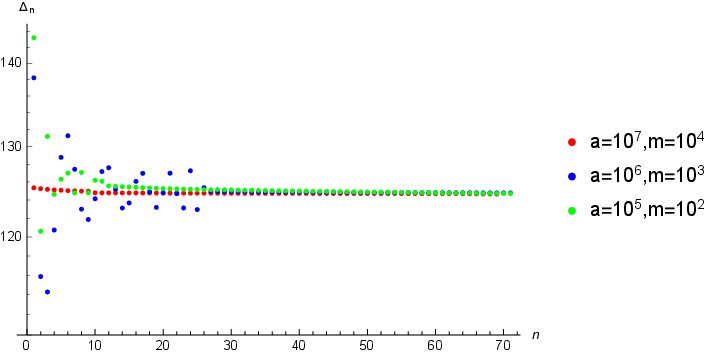}
\caption{For $\mathcal{C}=10^{-3}$,$\Delta_n$ converges to a asymptotic value. }
\label{fig4}
\end{figure}

In FIG.\ref{fig4}, by using (\ref{appro}), we plot $\Delta_n $ in terms of $n$, for different pairs of $(m,a)$ but with the same $\frac{m}{a}=10^{-3}$.
For large $n$, they tend to the same asymptotic value $\Delta_n=124.6$. This is in agreement with the WKB approximation, (\ref{treatment}), which results in $\Delta_n=\frac{\pi^{\frac{3}{2}}}{\sqrt{2}\sqrt{\mathcal{C}}}= 124.52$.

\begin{figure}[H]
\includegraphics[width=4.5in]{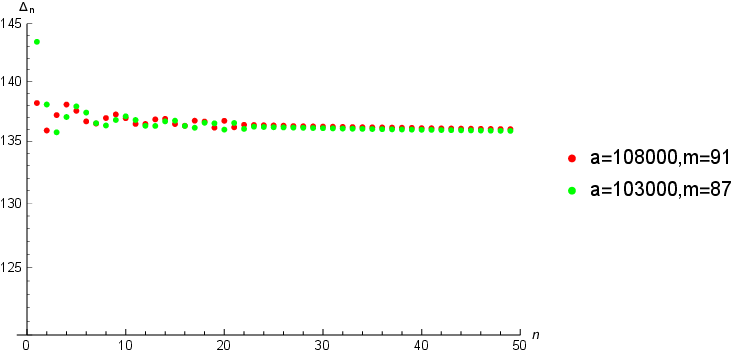}
\caption{For $\mathcal{C}=84\times 10^{-5}$,$\Delta_n$ converges to an asymptotic value, $\Delta_{\mathcal{C}}\simeq135.5$.}
\label{fig5}
\end{figure}

In FIG. \ref{fig5} asymptotic behaviour of $\Delta_n$ is shown for halos with  $\mathcal{C}=84\times 10^{-5}$.

For large couplings, scalarization by the surrounding DM halo with a given compactness gives rise to an asymptotic n-independent behavior: $\Delta_n\simeq 135.5$ which is compatible with the WKB approach that gives
$\Delta_n=\frac{\pi^{\frac{3}{2}}}{\sqrt{2}\sqrt{\mathcal{C}}}=135.86$.

This is similar to the result of \cite{Hod:2019pmb}, where for scalar Gauss-Bonnet model, $\Delta_n\equiv(\sqrt{\abs{\eta_{n+1}}}- \sqrt{\abs{\eta_n}})=\frac{\sqrt{3}}{2}\pi$ for $n\gg 1$ was reported.

\section{Conclusion}
We investigated spontaneous scalarization of a black hole immersed in a Hernquist dark matter halo within scalar-tensor theories featuring conformal coupling between the scalar field and dark matter. Our analysis demonstrates that the presence of a dark matter halo can trigger tachyonic instability in the scalar field, leading to nontrivial scalar configurations around black holes.
the scalarization phenomenon is governed solely by the Hernquist halo's compactness parameter $\mathcal{C}=\frac{m}{a}$, representing the ratio of the halo's total mass to its scale radius. Using the WKB approximation in the astrophysically relevant regime, and for large coupling constants, we derived an analytical quantization condition for the conformal coupling parameter,
which reveals that the coupling constant becomes discretized in terms of the inverse halo compactness:
$\eta_n=-\frac{(n+1/2)^2\pi^3}{2\mathcal{C}}$.
Our numerical results confirm that the discrete spectrum exhibits an asymptotic behavior $\Delta_n :=\sqrt{\abs{\eta_{n+1}}}- \sqrt{\abs{\eta_n}}=\frac{\pi^{\frac{3}{2}}}{\sqrt{2\mathcal{C}}}$ for large n, consistent with our approximation in the eikonal regime.  Our numerical results for various halo parameters confirm that as the compactness parameter decreases, the required coupling constants increase for fixed node numbers, and $\Delta_n$
converges to the predicted WKB asymptotic value independent of the specific values of $m$ and $a$.
This indicates that the halo compactness serves as the fundamental parameter characterizing DM-induced scalarization for the
Hernquist model.

\end{document}